\documentclass[]{mn2e}
\usepackage[dvips]{graphicx}
\usepackage{times}
\newcommand{\be}{\begin{equation}}
\newcommand{\ee}{\end{equation}}
\newcommand{\bea}{\begin{eqnarray}}
\newcommand{\eea}{\end{eqnarray}}
\begin{document}

\title[Planetary Migration and Resonant Exoplanets]{Planetary Migration and 
       Extrasolar Planets in the 2/1 Mean-Motion Resonance}
\author[C. Beaug\'e, S. Ferraz-Mello and T.A. Michtchenko]
       {C. Beaug\'e$^{1}$, S. Ferraz-Mello$^{2}$ and T.A. Michtchenko$^{2}$ \\
$^{1}$ Observatorio Astron\'omico, Universidad Nacional de C\'ordoba, Laprida 
       854, (X5000BGR) C\'ordoba, Argentina \\
$^{2}$ Instituto de Astronomia, Geof\'{\i}sica e Ci\^encias Atmosf\'ericas, 
       USP, Rua do Mat\~ao 1226, 05508-900 S\~ao Paulo,Brasil}

\date{}


\maketitle

\label{firstpage}

\begin{abstract}
We analyze the possible relationship between the current orbital elements fits
of known exoplanets in the 2/1 mean-motion resonance and the expected orbital 
configuration due to migration. We find that, as long as the orbital decay was 
sufficiently slow to be approximated by an adiabatic process, all captured 
planets should be in apsidal corotations. In other words, they should show a 
simultaneous libration of both the resonant angle and the difference in 
longitudes of pericenter. 

We present a complete set of corotational solutions for the 2/1 
commensurability, including previously known solutions and new results. 
Comparisons with observed exoplanets show that current orbital fits of three 
known planetary systems in this resonance are either consistent with apsidal 
corotations ({\it GJ876} and {\it HD82943}) or correspond to bodies with 
uncertain orbits ({\it HD160691}). 

Finally, we discuss the applicability of these results as a test for the 
planetary migration hypothesis itself. If all future systems in this
commensurability are found to be consistent with corotational solutions, then 
resonance capture of these bodies through planetary migration is a working 
hypothesis. Conversely, If any planetary pair is found in a different 
configuration, then either migration did not occur for those bodies, or it 
took a different form than currently believed.

\end{abstract}  


\section{Introduction}

It is well known that extrasolar planets are not where we imagined. Classical
planetary formation theories based on planetesimal accretion and 
core-instability for the giant planets predict bodies in quasi-circular orbits
and semimajor axes $a$ far from the star. For solar type stars, the minimum
semimajor axis is about 4 AU, which is the distance where non-rocky volatile 
elements can condense and accrete. However, many exoplanets do not follow this 
rule, but are found in highly eccentric orbits with $a < 1$ AU.

Two options have been proposed to explain this dilemma. In the first, it is
assumed that present cosmogonic theories are in fault, or at least they 
followed different routes in practically all other planetary systems (thus 
making our own Solar System a very particular case). In the second, exoplanets 
really did form far from the central star, but suffered a posterior decay in 
their semimajor axes towards their present sites. Thus was born what is 
usually referred to as the ``Hypothesis of Planetary Migration''. However, in 
order for migration to be a real theory and not just a simplistic escapade, 
two conditions must be met: {\it (i) } the existence of a plausible driving 
mechanism to explain the alleged decay in orbital energy, and {\it (ii)} 
concrete evidence that exoplanets did undergo such an evolution.

Two different driving mechanism have been presented in the last few years.
The first (Murray et al. 1998) is based on the interaction of the planets 
with a reminiscent planetesimal disk, and works in the same manner as 
migration of the giant planets in our Solar System (Fernandez and Ip 1984,
Hahn and Malhotra 1999). However, this mechanism does not seem to be 
sufficiently efficient. First, it requires a very large disk mass, of the 
order of 0.1 $M_{\odot}$, to explain an orbital decay of several astronomical 
units. Second, it is not completely evident that multi-planet systems should 
in fact undergo a simultaneous decrease in semimajor axis. Recall that in our 
system, Jupiter is believed to have suffered decrease in orbital energy, while 
Saturn, Uranus and Neptune have increased their value of $a$. 

The second proposed mechanism is the interaction of the planets with the 
gaseous disk. The existence of disk torques cause a transfer of energy and 
angular momentum from the planet to the gas and, if the disk parameters are 
chosen correctly, then the solid body should undergo a negative migration 
(i.e. ${\dot a}<0$). Several simulations have been performed in recent years 
(e.g. Snellgrove et al. 2001, Nelson and Papaloizou 2002, Kley 2003,
Papaloizou 2003) and these seem to indicate that the mechanism works 
reasonably well. However, certain aspects of this process are also 
problematical. For some disk parameters the migration can be positive, leading 
to an increase in semimajor axis, which is just the opposite desired result. 
In some other cases, the orbital energy may even exhibit random-walks with no 
secular variation. Nevertheless, it seems that this mechanism is the most 
probable process to explain migration.

Having found a plausible process for the orbital decay, we must now search 
for evidence that this really occurred in the exoplanets. This question is 
particularly important, since such large-scale {\it inward} migration did not 
happen in our case. A possible solution is to find a particular orbital 
characteristic of the extrasolar bodies, intimately related to migration, 
which can be used as (at least) indirect evidence of this process. Our Solar 
System presents two cases of confirmed migration (be it outward or inward): 
the giant planets and planetary satellites. As mentioned before, the outer 
planets migrated due to interaction with a remnant planetesimal disk, while 
many of the regular satellites of these same planets evolved due to tidal 
effects of the central mass. In this latter case, we know that an 
important consequence of the migration was capture of the satellites in exact 
mean-motion resonance (e.g. Colombo et al. 1974, Yoder 1979). A well known 
example is given by the Galilean satellites of Jupiter. It is known that these 
configurations cannot be explained solely with gravitational perturbations, 
but only through resonance trapping under the effects of an exterior 
non-conservative force. The case of our outer planets is different. It seems 
that they are not exactly in resonance due to certain random-walk 
characteristics of the driving mechanism itself (Hahn and Malhotra 1999). Thus,
we can conclude that although migration does not always lead to resonance 
trapping, the existence of massive bodies in exact mean-motion resonance can 
only be explained via a migration mechanism. 

The fact that our planets did not suffer a significant smooth inward migration
is consistent with the fact that none of them are trapped in resonance. What 
about the exoplanets? A good evidence in favor of migration would then be to 
analyze whether planetary systems do show mean-motion resonant relations. Of 
the 13 presently known planetary systems, including both confirmed and 
un-confirmed cases, we restrict ourselves to those systems where the ratio in 
semimajor axes is sufficiently small to assure significant gravitational 
interaction between the bodies. Choosing this limit to be $a_2/a_1 = 3$, we 
find that 6 systems satisfy this condition. These are: {\it GJ876}, 
{\it 55Cnc}, {\it 47Uma}, {\it HD82943}, {\it HD160691} and {\it Ups And}. Of 
these, five are believed to be in the near vicinity of mean-motion resonances, 
while {\it Ups And} is in a secular resonance. 

Although this proportion is very significant, it must be considered with care. 
Recent data analysis (Mayor et al., private communication) shows that the 
published orbit of {\it 47 Uma} may be questionable, and there is no general 
agreement if the system is near the 7/3 or 5/2 mean-motion resonances. Doubts 
also exist for {\it HD160691}, and the second planet is not yet confirmed.
In view of this debate, and considering the intrinsic errors in orbital fits,
the mere proximity of these systems to mean-motion resonances is not
evidence enough for migration, especially considering that there may be a 
natural tendency of researchers to place planets in commensurabilities even 
though they may not be near enough. For these reasons, we feel a more 
detail analysis is necessary.

This manuscript undertakes such a analysis. In Section 2 we present new 
results on the location and characteristics of corotations in the 2/1
resonance. Section 3 discusses the problem of planetary migration from the 
point of view of Adiabatic Invariant theory and we show that only 
corotational type configurations can apparently be expected in trapped 
planets. A comparison between these solutions and the current orbits of 
three exoplanetary systems are detailed in Section 4. Finally, conclusions 
close the paper in Section 5.

\section{General Apsidal Corotations for the 2/1 Mean-Motion Resonance}

Suppose two planets of masses $m_1$ and $m_2$ in coplanar orbits around a 
star of mass $M_0 \gg m_1,m_2$. Let $a_i$ denote the semimajor axis of the 
$i^{\rm th}$-planet ($i=1,2$), $e_i$ the eccentricity, $\lambda_i$ the mean 
longitude and $\varpi_i$ the longitude of the pericenter. All orbital elements 
correspond to Poincar\'e canonical relative coordinates (see Ferraz-Mello et 
al. 2004), which differ from the classical star-centered orbital elements in 
the second order of the planetary masses. We will suppose $a_1 < a_2$, thus 
the subscript 2 will correspond to the outer orbiting body. 

We now assume that both secondary masses are located in the vicinity of a 
resonance such that their mean motions $n_i$ satisfy the relation 
$n_1/n_2 \simeq (p+q)/p$. Both $p$ and $q$ are small integers and $q$ is 
usually referred to as the order of the resonance. The name 
``apsidal corotation'' (see Ferraz-Mello et al. 1993) is used to denote the 
simultaneous libration of both resonant angles:
\bea
\label{eq1}
\theta_1 &=& (p+q)\lambda_2 - p\lambda_1 - q\varpi_1 \\ 
\theta_2 &=& (p+q)\lambda_2 - p\lambda_1 - q\varpi_2 \nonumber .
\eea
It is straightforward to write 
$\theta_2-\theta_1 = q(\varpi_1 - \varpi_2) = q\Delta \varpi$, thus an apsidal 
corotation can also be identified with the libration of both $\theta_1$ and 
the difference in longitudes of pericenter. 

Once the short-period perturbations (associated with the synodic period) are
eliminated by an averaging process, the resulting system is a two degree of
freedom problem and can thus be specified by two angular variables, for 
example, $(\theta_1,\Delta\varpi)$. Their canonical conjugates are given by:
\bea
\label{eq2}
I_2 &=& {1 \over q} L_2 \biggl(1 - \sqrt{1-e_2^2}\biggr) \\
I_1 &=& I_2 + {1 \over q} L_1 \biggl(1 - \sqrt{1-e_1^2}\biggr) \nonumber .
\eea
The quantity $L_i = m'_i \sqrt{\mu_i a_i}$ is the modified Delaunay momenta
related to the semimajor axis in Poincar\'e variables (see Laskar 1991, 
Ferraz-Mello et al. 2004), $\mu_i = G (M_0 + m_i)$, and $G$ is the 
gravitational constant. The factor $m'_i$ is a reduced mass of each body, 
given by:
\be
\label{eq3}
m'_i = {m_i M_0 \over m_i + M_0}  .
\ee
It is easy to see (e.g. Michtchenko and Ferraz-Mello 2001) that the total
planar angular momentum of the system, in itself an integral of motion, is 
given by:
\be
\label{eq4}
J_{\rm tot} = L_1 + L_2 - I_1 .
\ee
Similarly, the complete averaged Hamiltonian of the system can be expressed 
in terms of the orbital elements as:
\be
\label{eq5}
F = -\sum_{i=1}^2 {\mu_i^2 {m'}_i^3 \over 2 L_i^2}
 - F_1(m_1,m_2,a_1,a_2,e_1,e_2,\theta_1,\Delta\varpi) ,
\ee
where the disturbing function $F_1$ denotes the gravitational interaction
between both planets. Further details can be found in Beaug\'e and Michtchenko 
(2003). With this in mind, apsidal corotations can now be constructed from the 
conditions: 
\bea
\label{eq6}
{dI_1 \over dt} &=& 0 \hspace*{0.5cm} ; \hspace*{0.8cm} 
{dI_2 \over dt} = 0 \\
{d\theta_1 \over dt} &=& 0 \hspace*{0.5cm} ; \hspace*{0.5cm} 
{d\Delta\varpi \over dt} = 0 \nonumber .
\eea
In other words, they are fixed points of the averaged Hamiltonian (\ref{eq5}). 
Once the short-period variations are re-introduced, corotations actually 
correspond to periodic orbits (Hadjidemetriou 2002, Hadjidemetriou and 
Psychoyos 2003). 

It is important to emphasize two points. First, apsidal corotations are
zero-amplitude solutions in the averaged problem. In a real system, the planets
may in fact undergo finite-amplitude oscillations around these points, thus
describing quasi-periodic orbits in real space. The simulations by Lee and
Peale (2002) of the {\it GJ876} show such a behavior. A second, and very 
important note, is that it is possible that the amplitude of oscillation in 
$\Delta \varpi$ be sufficiently large to show an actual circulation (and not a 
libration) of this angle. In other words, the difference in longitudes of
pericenter may vary from zero to 360 degrees, although topologically the
solution is still an apsidal corotation. Thus, a finite-amplitude corotation 
is really defined in terms of a separatrix crossing to a different mode of 
oscillation not present in the unperturbed dynamical system, and not solely in 
terms of the temporal variation of $\Delta \varpi$.

In a recent study, Beaug\'e et al. (2003) (hereafter referred to as BFM2003)
and Ferraz-Mello et al. (2003) performed a systematic search for different
types of corotational solutions in the 2/1 and 3/1 resonances. Among our
first results we found that, up to second order of the masses, apsidal 
corotations only depend on the real masses of the planets through the ratio 
$m_2/m_1$. In other words, these solutions are not function of the individual 
values, and are thus independent of the inclination of the orbital plane of 
the planets with respect to the observer (as long as both planets share the 
same plane). This is a very interesting point, since it allows us to bypass 
the limitations in Doppler orbital fits. Secondly, we also found that these 
periodic orbits only depend on the semimajor axes through $a_1/a_2$. Since 
this ratio is only an indication of the proximity to exact resonance, it is 
independent of the individual values of the semimajor axes themselves. As a 
consequence of these properties, we were able to obtain the apsidal corotations
as level curves of $\theta_1$, $\Delta \varpi$ and $m_2/m_1$ in the plane of 
eccentricities $(e_1,e_2)$, and these results were seen to be extremely 
general. They are valid for any planetary system, independently of the values 
of the real masses and the distance from the central star.

For the 2/1 resonance, our results showed the existence of three types
of corotational solutions. Aligned apsidal corotations are characterized by
equilibrium values of the angles equal to $(\theta_1,\Delta \varpi)=(0,0)$.
Anti-aligned solutions are given by $(\theta_1,\Delta \varpi)=(0,\pi)$. Both
these families were previously known by other authors (e.g. Lee and Peale 2002,
Hadjidemetriou 2002). However, we also discovered a new type of orbits, called
asymmetric apsidal corotations, which were characterized by values of 
$(\theta_1,\Delta \varpi)$ different from zero or $\pi$ (see Greenberg 1987
for similar results for the Galilean satellites). Finally, to each value of 
$(e_1,e_2)$ there seemed to correspond only one equilibrium value of the mass 
ratio $m_2/m_1$. Similar results were also noted for the 3/1 resonance.

Due to the inherent limitations of our model, we were only able to detect
apsidal corotations with eccentricities up to $e_i = 0.5$. Moreover, we only 
analyzed cases where $m_2/m_1>0.1$. This limit was chosen simply for 
computational reasons. Recently, however, two new results came to our 
attention. Numerical studies by Hadjidemetriou and Psychoyos (2003) noted that 
some eccentricities allowed for two different equilibrium values of $m_2/m_1$, 
one of them smaller than $0.1$. Secondly, the same authors also found a new 
type of corotational orbit, characterized by 
$(\theta_1,\Delta \varpi)=(\pi,\pi)$ for very high values of $e_1,e_2$. 

However, both these studies only analyze a restricted number of initial 
conditions and do not present general results in the $(e_1,e_2)$ plane.
Thus, a more general analysis may be useful. Since our previous model is not
sufficiently adequate for such high eccentricities, for the present paper we 
adopted a new semi-analytical approach based on the so-called Extended
Schubart Averaging (Moons 1994), where the Hamiltonian (\ref{eq5}) and 
conditions (\ref{eq6}) are solved numerically. We also lifted the restriction 
in the minimum mass ratio, thus allowing us to study the general corotational 
solutions with no restriction whatsoever in their parameters. 

\begin{figure}
\centerline{\includegraphics*[width=20pc]{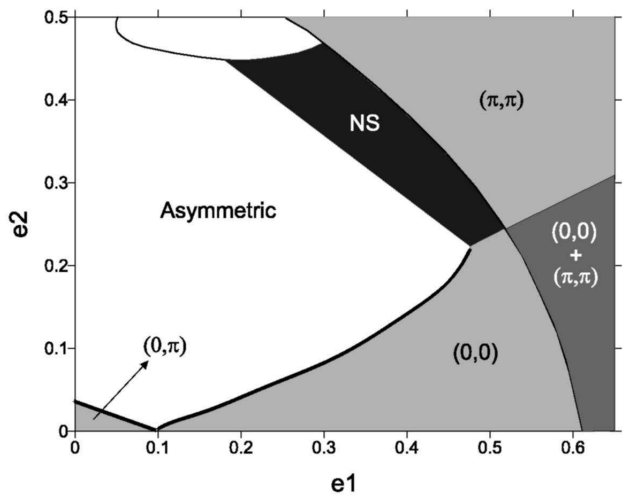}}
\caption{Domains of different types of corotational solutions in the 2/1
mean-motion resonance, as seen in the plane of orbital eccentricities of
both planets. NS is the region with no stable solutions. See text for further
explanations.}
\label{fig1}
\end{figure}

Figure \ref{fig1} shows the plane of eccentricities for the 2/1 resonance,
where we have drawn the limits of the domains of all types of solutions. Each
is marked by the equilibrium values of the angles, except the asymmetric 
region, plus a new domain denoted by NS (i.e. No Solution). In this region 
there are no stable apsidal corotations for any values of the mass ratio. This 
is due to close encounters, where the potentially equilibrium values of the 
angles correspond to quasi-collisions between the planets. We will return to 
this point further on. Note that the $(\pi,\pi)$-corotations are located 
beyond the collision curve, at very high eccentricities; nevertheless this 
domain intersects the aligned apsidal corotation regime for relatively low 
values of $e_2$. In this intersection, two distinct types of stable solutions 
(aligned and $(\pi,\pi)$) coexist, although for different values of the masses.
Finally, the asymmetric zone is actually divided into two distinct regions 
which are not connected via a smooth variation of the planetary masses.

\begin{figure}
\centerline{\includegraphics*[width=20pc]{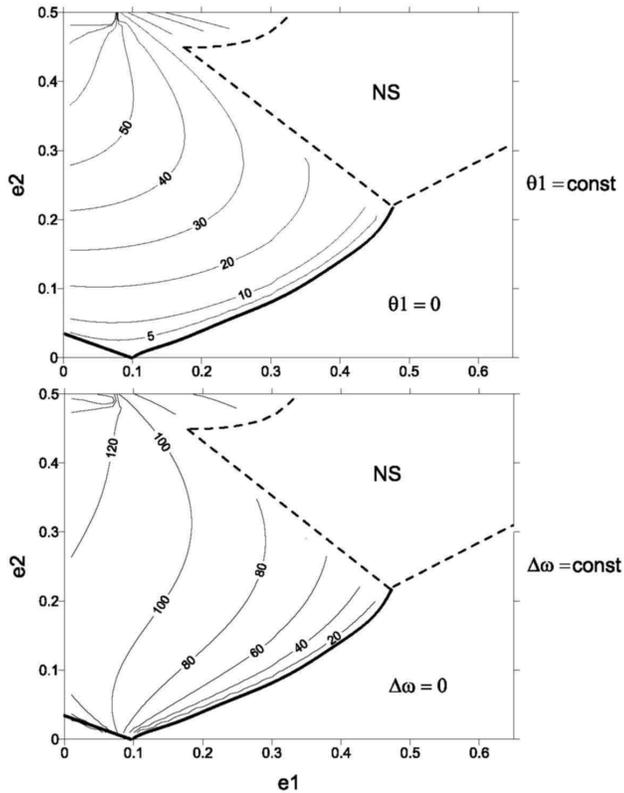}}
\caption{Top: Level curves of equilibrium values of the resonant angle
$\theta_1$ for all corotations below the collision curve. Bottom: idem, but for
the values of $\Delta \varpi$.}
\label{fig2}
\end{figure}

The equilibrium values of the resonant angle $\theta_1$ and of the difference
in pericenter $\Delta \varpi$ are shown in Figure \ref{fig2}. All types of 
solutions are shown with the exception of the $(\pi,\pi)$-corotations, which
will be discussed further on. This figure is analogous to the results 
presented in BFM2003, although extended to higher eccentricities. We note 
two main differences. First, asymmetric apsidal corotations are limited to 
$e_1 < 0.5$, after which only aligned orbits are possible. Second, the NS
region was not detected in our previous work. This was due to the limitations 
of the analytical model, where our expansion of the Hamiltonian underestimated 
the values of the function in the vicinity of the collision curve. 
Nevertheless, all the solutions with $e_1+e_2 \le 0.5$ are equal in both 
determinations.

\begin{figure}
\centerline{\includegraphics*[width=20pc]{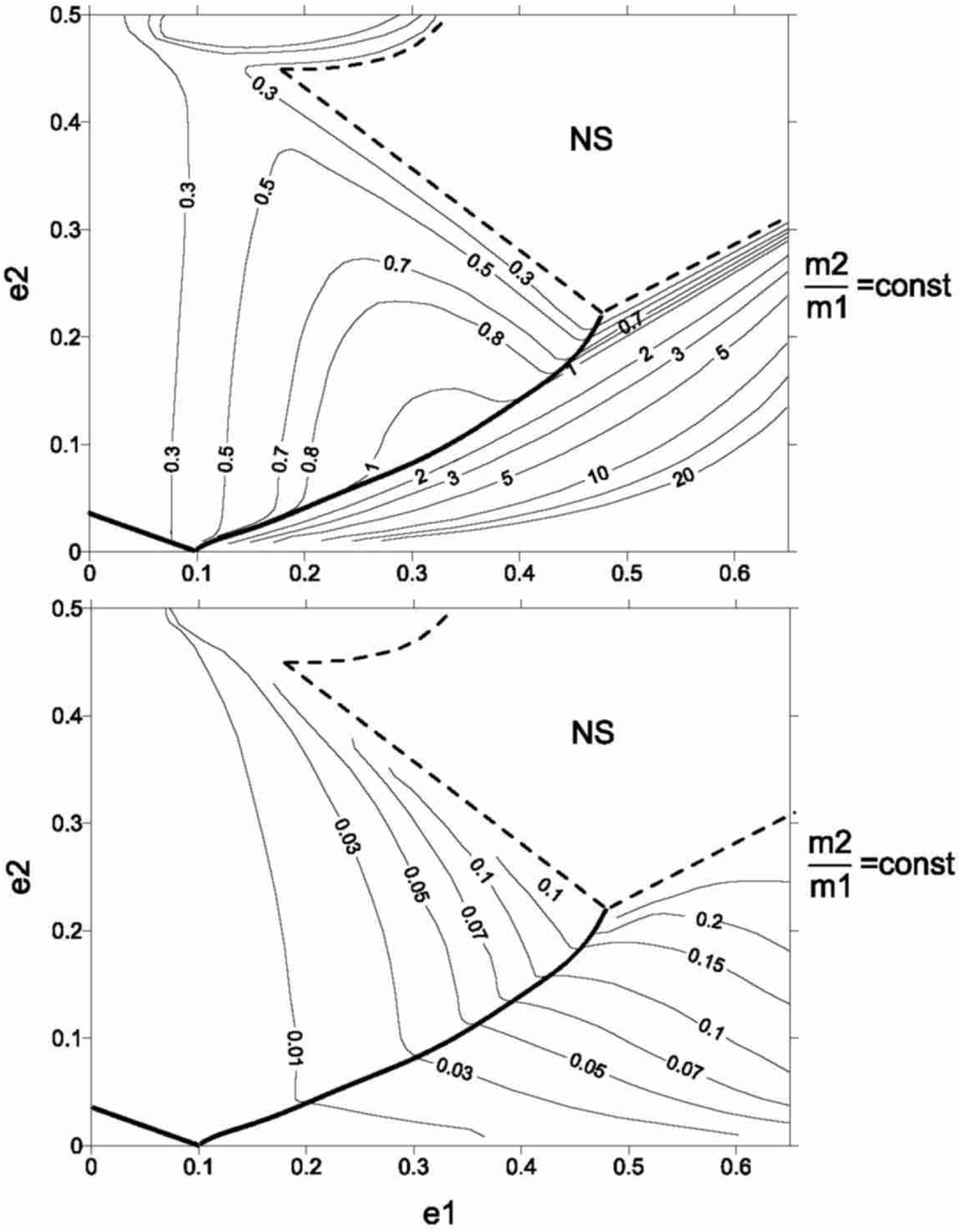}}
\caption{Level curves of constant mass ratios $m_2/m_1$ for stable corotations
in the 2/1 resonance. Note that most values of the eccentricities have two
possible solutions. Larger values of the masses are plotted on the top graph,
while smaller values are shown on the bottom. }
\label{fig3}
\end{figure}

\begin{figure}
\centerline{\includegraphics*[width=20pc]{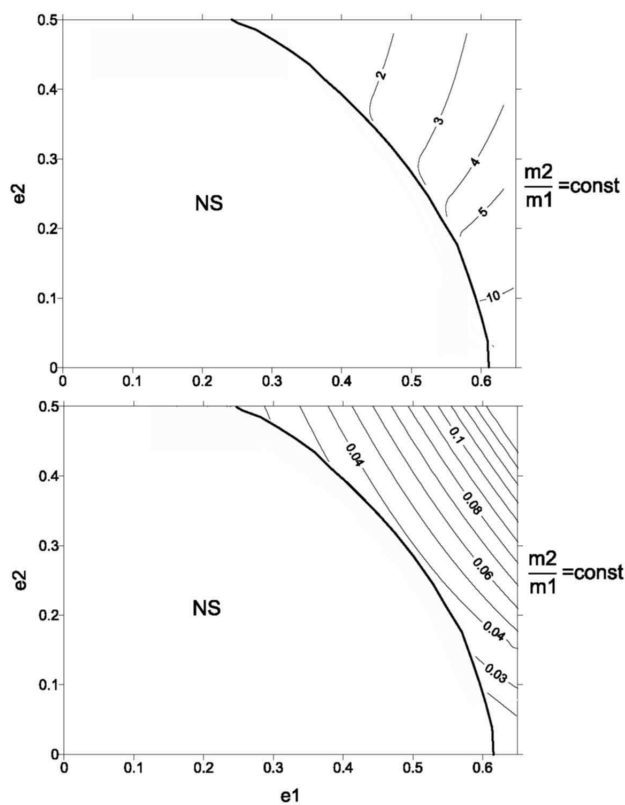}}
\caption{Idem previous figure, but for the $(\pi,\pi)$-corotations.}
\label{fig4}
\end{figure}

The level curves of mass ratios for $(0,0)$, $(0,\pi)$ and asymmetric 
solutions are shown in Figure \ref{fig3}.  Notice that for practically all
eccentricities there are two distinct equilibrium values of $m_2/m_1$, in
accordance with the predictions of Hadjidemetriou and Psychoyos (2003). Larger
ratios are shown on the top graph, while smaller quantities on the bottom plot.
Figure \ref{fig4} shows similar results, now for the $(\pi,\pi)$-corotations. 

\begin{figure}
\centerline{\includegraphics*[width=20pc]{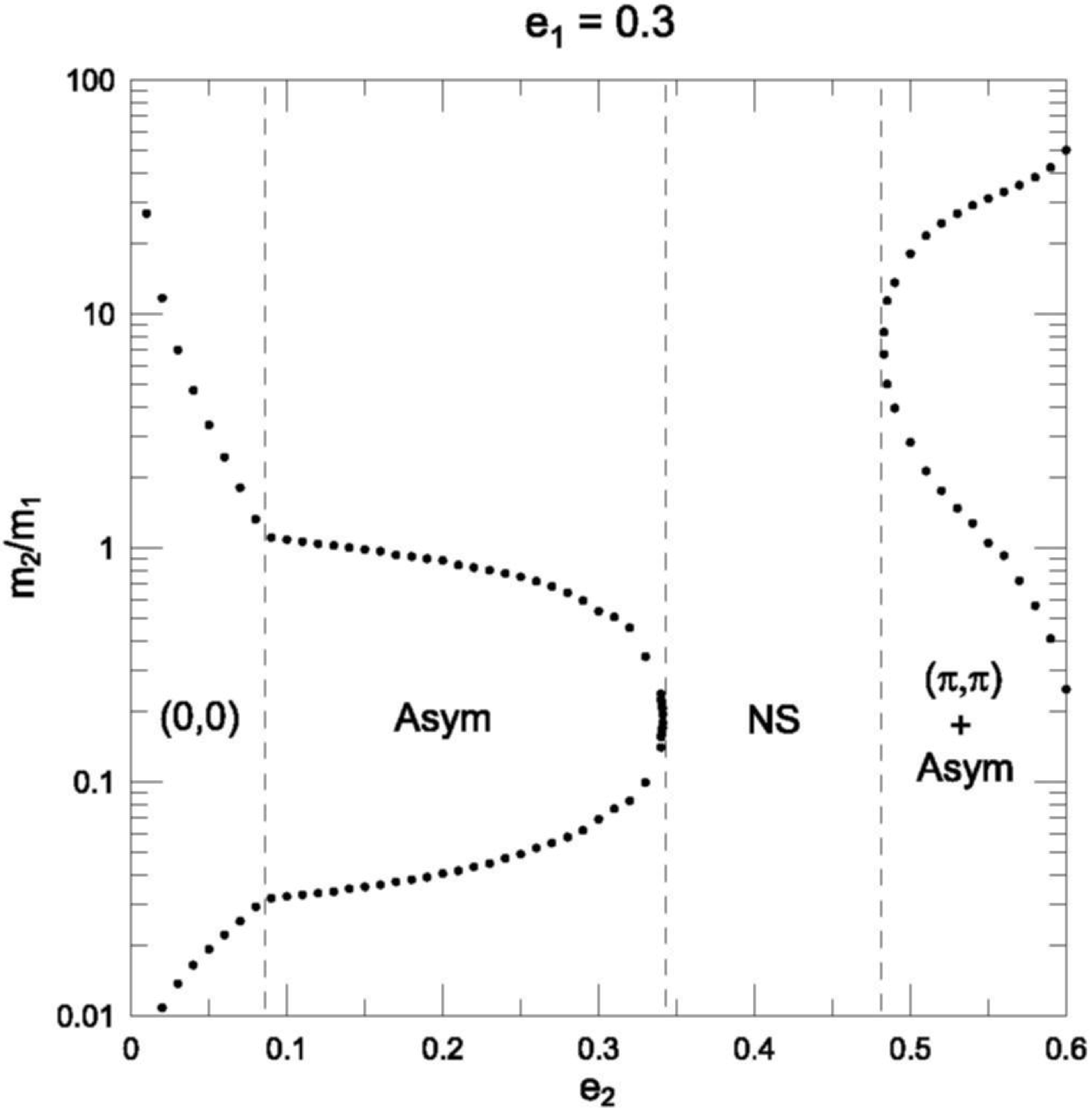}}
\caption{Mass ratio of corotations, as function of the eccentricity of the 
outer body, for two values of $e_1$. Red shows $e_1=0.3$ while black 
corresponds to $e_1=0.5$. Notice that the region of No Solutions appears when
both mass ratios per eccentricity fold into a single solution.}
\label{fig5}
\end{figure}

In order to have a better understanding of the origin of the NS region in
Figure \ref{fig1}, we studied the variation in the mass ratios as
function of $e_2$ for a constant value of the eccentricity of the inner
planet ($e_1=0.3$). This value was chosen so that it intersects the domain of 
No Solutions, as well as aligned and asymmetric apsidal corotations. Results 
are shown in Figure \ref{fig5}, where we can see that the boundaries of the NS 
region are characterized by a coalescence of two values of the mass ratios. 

A final data we wish to present at this point is the period of motion of 
oscillations around apsidal corotations. As early as the analysis of 
{\it GJ876} by Lee and Peale (2002), it is known that exoplanets do not 
necessarily have to be in an exact periodic orbit (i.e. zero-amplitude apsidal
corotation), but can exhibit a finite amplitude oscillation around this 
solution with a certain period. This quantity, at least for the linear 
approximation, can be determined calculating the Hessian of the Hamiltonian 
evaluated at each apsidal corotation. However, due to the order of the 
characteristic equation, this method is very time-consuming and sensitive to 
numerical errors. For these reasons, in the present work we employed a more 
numerical approach. 

\begin{figure}
\centerline{\includegraphics*[width=20pc]{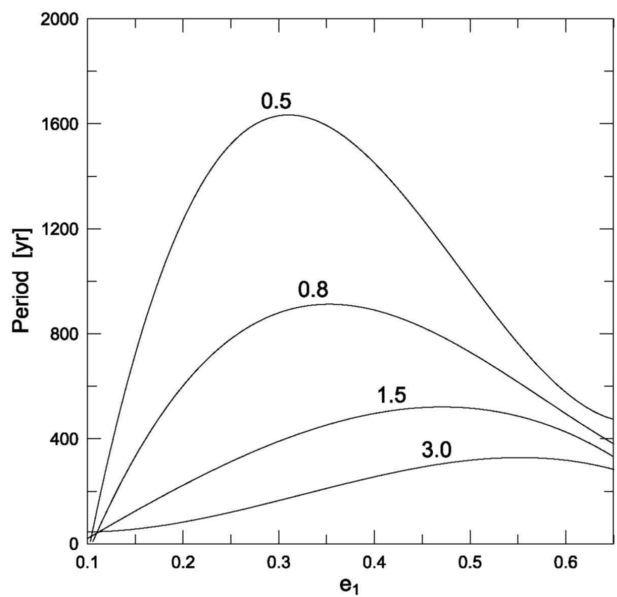}}
\caption{Period of infinitesimal oscillations around corotations, for four 
different mass ratios, as function of the eccentricity of the inner planet.}
\label{fig6}
\end{figure}

Considering a fixed mass ratio, we first performed numerical simulations of 
the evolution of the system along the family of periodic orbits, in a manner 
analogous as presented in Ferraz-Mello et al. (2003). We then calculated a
running Fourier analysis of the angular variables at given times, and 
calculated the period $\tau$ associated with the largest amplitude. 
Simultaneously, we also estimated the averaged planetary eccentricities, thus 
obtaining a relation between $\tau$ and $e_1$. Results are presented in Figure 
\ref{fig6}, in units of years, for four different mass ratios. These periods 
correspond to $a_2=1$ AU and $m_1 = M_{\rm Jup}$. It must be noted that the 
curves have been smoothed, both to eliminate spurious differences between
adjacent points, and to soften the separatrix separating symmetric from 
asymmetric solutions. Thus, individual values must be considered more 
qualitative than quantitatively correct, although the general trend is
fairly accurate.

Even with these notes of caution in mind, the plot still gives valuable 
information. We can see that the period of oscillation increases for smaller 
values of the mass ratio, and the rate is practically inverse-linear. Thus, 
the maximum $\tau$ for $m_2/m_1 = 3$ (similar to the {\it GJ876} system) is 
about 300 years, while for a mass ratio of 0.5, the maximum period is about 
six times larger. Second, comparing these results with Figure \ref{fig3}, we 
note that asymmetric apsidal corotations have much larger periods than 
symmetric solutions. This characteristic will prove important in later 
sections. 

Finally, as mentioned before, the quantities in the graph correspond to 
$a_2=1$ AU and $m_1 = M_{\rm Jup}$. For other values of these parameters, the 
resulting period must be scaled according to:
\be
\label{eq6bis}
\tau = \tau_0 {a_2}^{3/2} \biggl( { m_1 \over M_{\rm Jup}} \biggr)^{-1} 
\ee
where $\tau_0$ is the corotational period given in the figure. Thus, although
the position of the apsidal corotations are only function of $m_2/m_1$, the 
periods of oscillation are linearly dependent on the values of the individual 
masses.

\section{Planetary Migration and the Adiabatic Invariant Theory}

In principle, the results presented in the previous section should constitute
a catalog of all corotational solutions in the 2/1 resonance. If all extrasolar
planets in this resonance lie in apsidal corotations, then their orbits should 
be well represented in those figures. However, what evidence do we have that 
all exoplanets in the 2/1 commensurability are in fact in apsidal corotation? 
All dynamical analysis of the {\it GJ876} system predict such a configuration, 
so there is little doubt that these planets satisfy our assumption. However, it
is not immediate that the same should be generally valid for all other systems.

If all resonant exoplanetary systems acquired there present orbits as a result 
of planetary migration, then an important test would be to check whether 
captured migrating bodies do exhibit apsidal corotations. Recent hydrodynamical
simulations (e.g. Snellgrove et al. 2001, Kley 2003, Papaloizou 2003, etc.)) 
of the evolution of two planets immersed in a gaseous disk, have always shown 
corotational final orbits. Kley (2003), modeled {\it 55 Cnc}, and capture 
occurred in the 3/1 commensurability. In the other two papers the simulated 
system was {\it GJ876} and trapping occurred in the 2/1 mean-motion resonance. 
Other works, such as Nelson and Papalozoiu (2002) have included a modeled 
migration in the equations of motion of the planets, as constant perturbations 
in the angular momentum and orbital energy. Solving these equations 
numerically, they have also found corotations as a final result.

\subsection{Numerical Simulations of Resonance Capture}

In order to test whether these results are only valid for point values of
disk parameters or even for certain types of driving mechanisms, we performed
a series of numerical simulations of the planetary migration. In this series,
we studied the trapping process and posterior evolution of the system inside 
the resonance (general three-body problem) for a wide range of exterior 
non-conservative forces. Each force was modeled as an additional term added to 
Newton's equations of motion, and all runs simulated capture in the 2/1 
mean-motion resonance. We adopted various types of forces, including: 
{\it (i)} Tidal interactions (Mignard 1981), {\it (ii)} interaction with a 
planetesimal disk (modeled according to Malhotra 1995) and, {\it (iii)} disk 
torques modeled similar to Nelson and Papalozoiu (2002). Most of these 
mechanisms can be reproduced as particular cases of a Stokes-type 
non-conservative force of the type:
\be
\label{eq7}
{d^2{\bf r} \over dt^2} = -C({\bf v} - \alpha {\bf v}_c)
\ee
where ${\bf r}$ is the position vector of the body (reference frame centered 
in the star), ${\bf v}$ is its velocity vector and ${\bf v}_c$ is the circular 
velocity vector at the same point. Unlike usual Stokes drag where $\alpha$ is
fixed by the characteristics of the gas, in this generic case both coefficients
$C$ and $\alpha$ can be taken as external parameters and varied in each run. 
The first is usually considered defined positive (i.e. $C > 0$) while the
second can take any value. From Beaug\'e and Ferraz-Mello (1993) and Gomes 
(1995) it can be seen that, to first order and in the absence of additional 
gravitational perturbations, the effects of the force (\ref{eq7}) in the 
semimajor axis and eccentricity are given by:
\be
\label{eq8}
a(t) = a_0 \exp{(-At)} \hspace{0.5in} ; \hspace{0.5in} 
e(t) = e_0 \exp{(-Et)}
\ee
where $a_0$ and $e_0$ are the initial conditions at $t=0$, and $A$,$E$ are the
inverse of the e-folding times in each orbital element. These quantities are 
given by:
\be
\label{eq9}
A = 2C(1-\alpha) \hspace{0.5in} ; \hspace{0.5in} 
E = C\alpha .
\ee
Thus, $\alpha=1$ gives a non-conservative force that gives an exponential
decrease in semimajor axis but no change in the eccentricity, analogous to 
Malhotra's (1995) model of planet-planetesimal interactions. Moreover, when 
$\alpha<0$ the force acts to increase the value of the eccentricity, and the 
opposite occurs when $\alpha>0$. This can be used to model different types of
behavior noted in planet-disk interactions, depending on the preponderance of
external Limblad or co-orbital resonances (see Goldreich and Sari 2003). 

Finally, we can consider the e-folding times as input parameters of the 
simulation and deduce the coefficients accordingly:
\be
\label{eq10}
C = {1 \over 2} A + E \hspace{0.5in} ; \hspace{0.5in} 
\alpha = {E \over C}.
\ee
With all these options we hope to have a fairly general idea of the capture 
process in the 2/1 resonance under a variety of conditions and physical models.
Of course this list is not complete and it is not our intention to model all 
possible interactions, but it does give a general idea of the type of behaviors
that can be expected.

\subsection{Corotational Families as Evolutionary Tracks}

Using these models for the driving mechanism, we performed a series of
numerical simulations of the evolution (and resonance capture) of two planets 
with a given mass ratio, and initial circular orbits with $a_1=5.2$ AU 
and $a_1=8.5$ AU. These semimajor axes place the bodies outside, but close to,
the 2/1 mean-motion resonance. In particular, we did several runs with a
non-conservative force given by equation (\ref{eq7}), and adopting different 
values of the e-folding times in the range $A \in [10^{-7},10^{-4}]$ and 
$E \in [10^{-11},10^{-4}]$. 

In accordance with other works, all our runs ended in apsidal corotations. 
Although dynamical $\theta$-librations have previously been detected in 
resonance trapping for the restricted three-body problem with $m_2=0$ (see 
Beaug\'e and Ferraz-Mello 1993), it seems not likely (or not possible) for the 
general planetary case in the 2/1 commensurability. Even though we cannot rule 
out their existence, if present they should have small trapping probabilities 
or be the consequence of very particular kind of dissipation forces. Perhaps 
future analytical work will solve this question categorically; however, and 
for the remainder of this work, we will assume that only apsidal corotations 
are available for trapped bodies.

\begin{figure}
\centerline{\includegraphics*[width=20pc]{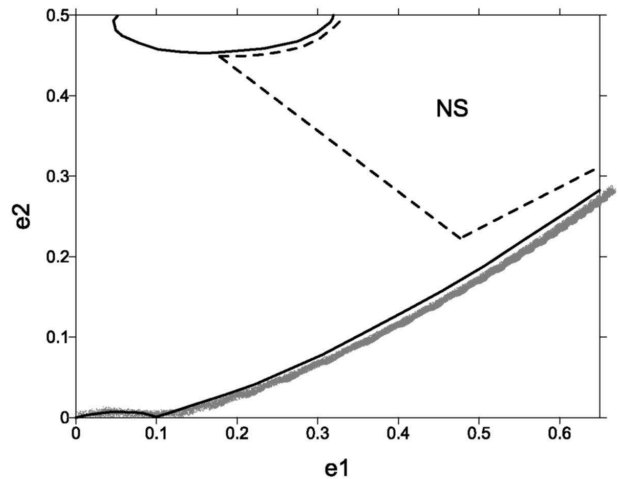}}
\caption{Relation between eccentricities of inner and outer planets during the 
orbital evolution inside the 2/1 resonance, using $m_2/m_1=1.5$. Grey symbols
show the results of several numerical simulations with different values of 
$A \in [10^{-6},10^{-4}]$ and $E \in [10^{-11},10^{-4}]$. Black lines show the 
two families of zero-amplitude corotations for this mass ratio: asymmetric 
solutions for high values of $e_2$ and aligned orbital configurations in the
other case.}
\label{fig7}
\end{figure}

Typical results (using $m_2/m_1 = 1.5$) are shown in Figure \ref{fig7}, where 
we have plotted the relationship between the eccentricities of the bodies 
prior to capture and during the orbital evolution inside the commensurability. 
The results of {\it all} numerical simulations are shown in gray symbols. 
Although different driving mechanisms may yield solutions which vary in 
capture timescales or amplitudes of libration, we can see that all points fall 
in the same region of the plane $(e_1,e_2)$. In fact, since the initial orbits 
were circular, the gray symbols define an ``evolutionary curve'' of the 
system, in which the eccentricities evolve from the origin to the right-hand 
side of the graph as function of time. Some values of $A,E$ yield equilibrium 
eccentricities (see Lee and Peale 2002), in which case the evolution stops at 
some critical value of $(e_1,e_2)$. Conversely, for other values of the 
e-folding times, evolution continues until $e_1$ reaches quasi-parabolic 
values and both planets collide.

In the same figure we have also plotted the region of No Solutions and, in 
black continuous lines, the families of corotations for this mass ratio. Three
families exist: asymmetric solutions lie on the top of the graph, for high 
values of $e_2$ and low values of $e_1$. Symmetric (aligned and anti-aligned) 
apsidal corotations lie in the lower half of the plot. Each individual solution
of a given family is parametrized by different values of the total angular 
momentum. Note that the symmetric family shows a very good agreement with the 
numerical simulations of the evolution of the planets. This in fact shows that 
during the capture process, the system evolves adiabatically following the 
stable equilibrium solutions of the conservative system. Thus, the family of 
apsidal corotations does not only point the present possible locations of 
extrasolar planetary systems in the vicinity of the 2/1 resonance, but can 
also give information about the routes the bodies took from initially 
quasi-circular orbits towards their present locations. 

This interpretation is possible as long as the driving mechanism of the 
migration is sufficiently slow, compared to the characteristic timescale of
the conservative perturbations, so that the system can be well approximated
by an smooth-varying Hamiltonian system. The Adiabatic Invariant Theory
(e.g. Neishtadt 1975, Henrard 1982) shows that this is satisfied if the
ratio between the period of oscillation around the apsidal corotation and the 
time derivative of the semimajor axis be much smaller than unity. In other 
words,
\be
\label{eq11}
\varepsilon \equiv \tau A \ll 1.
\ee
In order to quantify this relation, let us recall equation (\ref{eq6bis}), 
consider Jupiter-size planets and concentrate on the maximum values of $\tau$.
For initial semimajor axis $a_2$ in the vicinity of present day Jupiter, these
results seem to indicate that adiabaticity is satisfied if the migration
time-scale $1/A \gg 5 \times 10^3$ years for $m_2/m_1=0.5$ and 
$1/A \gg 8 \times 10^2$ years for $m_2/m_1=3$. For much smaller semimajor axes 
(e.g. $a_2 = 0.3$ AU), these numbers fall to values of the order of $10^2$ 
years. Thus, any dissipative force with migration timescale much larger than
this should be adiabatic and thus its evolution well modeled by the families
of corotational solutions. 

To date there is no concrete evidence as to the real duration of the migration,
although it is believed to have taken between $10^5$ and $10^7$ years 
(see Trilling et al. 2002). If this is indeed the case, then the adiabatic 
approximation should be a good model, at least close to the central star. 
Even if these limits are too conservative and an extremely fast Type I 
migration dominated the evolution, the mechanism most probably had a smooth 
decay in magnitude with time, thus becoming much slower towards the end of the 
process. From some point on then, the mechanism should satisfy condition 
(\ref{eq11}). If true, then it seems that any migration process should lead 
to present orbital configurations consistent with apsidal corotations. 

\begin{figure}
\centerline{\includegraphics*[width=20pc]{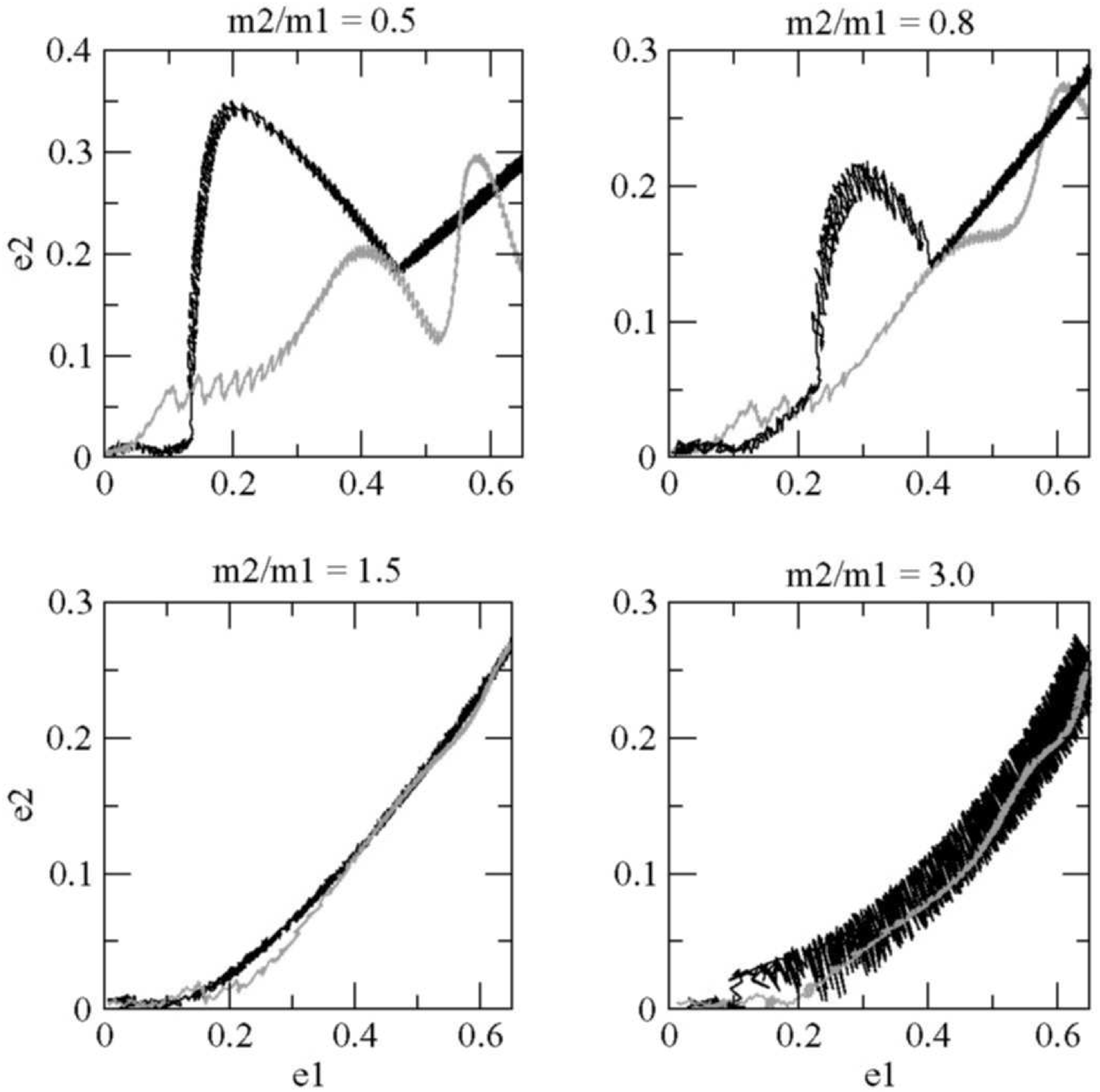}}
\caption{Numerical simulations of adiabatic migration (black symbols) and
non-adiabatic (gray) for four different mass rations. All plots show the
evolutionary tracks in the eccentricities plane.}
\label{fig8}
\end{figure}

In order to test this idea, Figure \ref{fig8} shows, for each mass ratio 
discussed in the previous paragraphs, two simulations of the capture process, 
one with $A = 10^{-6}$ (black) and the other considering  an extremely rapid
migration: $A = 10^{-4}$ (gray). Results are shown in the eccentricity plane 
in both cases. For the two larger mass ratios both simulations follow 
practically the same routes, and are consistent with the corotational families.
Recall that these mass ratios have small periods of oscillation and no 
asymmetric apsidal corotations. The top graphs show a different story. The 
system with $m_2/m_1 = 0.8$ shows fair agreement between both simulations for 
symmetric apsidal corotations, but completely different results for the 
asymmetric region. The results for $m_2/m_1 = 0.5$ are an extreme case. The 
fastest migration shows very little in common with the adiabatic evolution, 
although capture still takes place and both eccentricities continue to grow as 
function of time. 

\begin{figure}
\centerline{\includegraphics*[width=20pc]{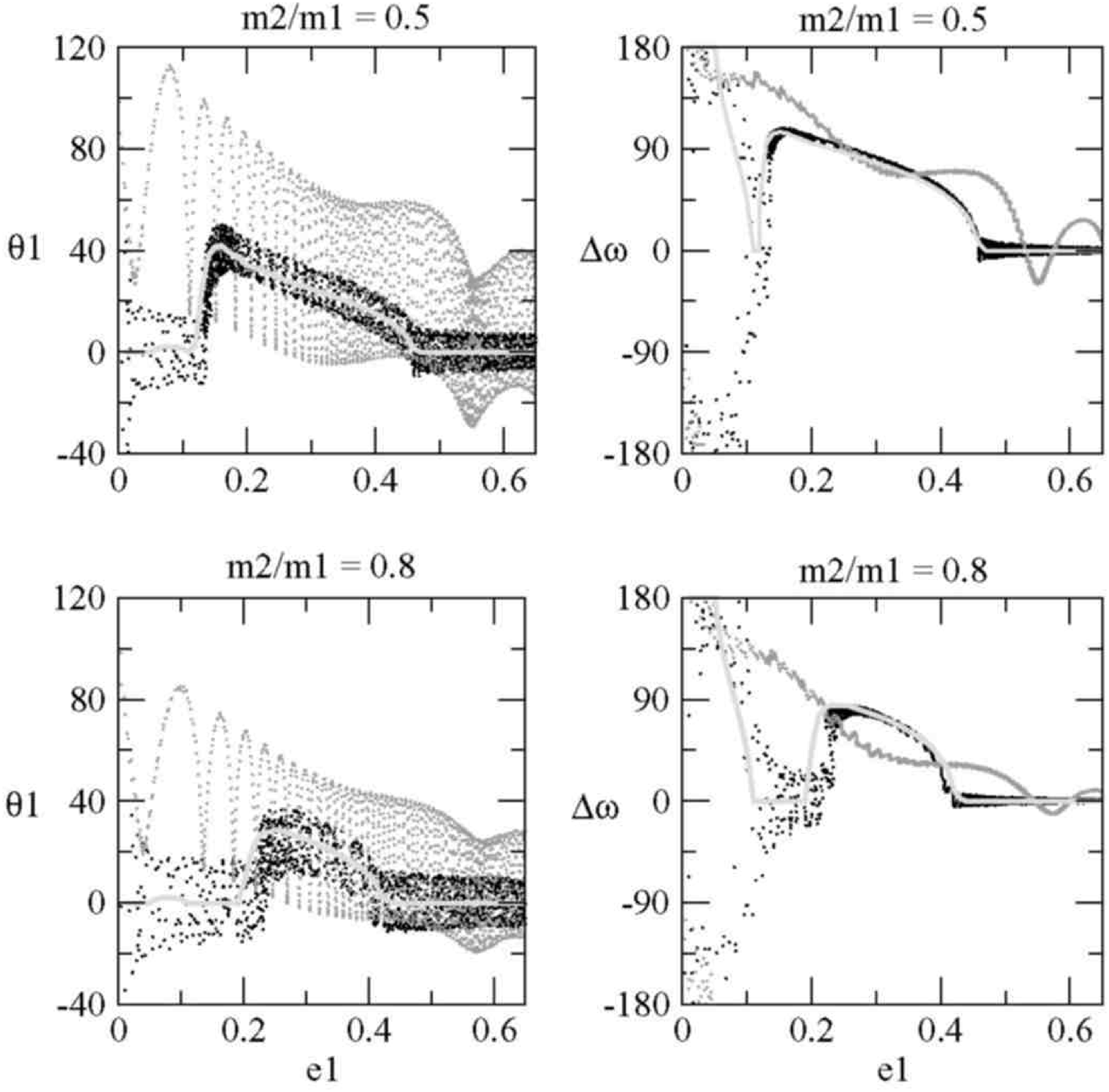}}
\caption{Same as before for only two mass-ratios, and showing the temporal 
evolution of the resonant angle $\theta_1$ (left) and difference in longitudes
of pericenter (right). The light-gray continuous lines show the analytical
corotational solutions parametrized by the eccentricity of the inner planet.}
\label{fig9}
\end{figure}

Figure \ref{fig9} presents the evolution of both angular variables, as
function of the growing $e_1$, for the two smallest mass ratios. Colors are 
the same as the previous figure, with black lines corresponding to the slowest
(adiabatic) migration and gray to the fastest. We can see that a non-adiabatic 
force not only implies different evolutionary tracks in the eccentricity plane,
but also in the angular variables. Interestingly, in both cases the largest
dissipation causes very evident asymmetric apsidal corotations which have no 
association with the conservative equilibrium solutions. This seems to 
indicate that, perhaps, the lack of adiabaticity is also accompanied by a 
change in the equilibrium solutions. Only for high values of $e_1$, accompanied
by small values of the semimajor axes (due to the orbital decay) do both 
curves reasonably agree, consistent with a decrease in $\tau$ compared with 
the migration timescale.

\section{Confirmed and Unconfirmed Planetary Systems in the 2/1 Resonance}

From the previous simulations we can conclude that for mass ratios larger
than 1.5, even a fast planetary migration leads to evolutionary tracks
consistent with our corotational families. We can then proceed to test this 
idea with the known planetary systems in the vicinity of the 2/1 resonance, be 
them confirmed or unconfirmed bodies. Recent stability analysis of several 
resonant exoplanets have shown that in some cases (e.g. {\it 47 Uma}) a 
corotation is not the only dynamically stable configuration. In fact, apart 
from the well-studied case of {\it GJ876}, it is not absolutely certain 
whether any other real system is in an actual apsidal corotation. However, as 
deduced from the analysis of the previous section, planetary migration does 
seem to imply this type of solutions. 

In view of this, a good test for the migration hypothesis is to check whether 
current orbital fits are {\it consistent} with apsidal corotations. If they are
not, then we stand with two possibilities. Perhaps the orbital fits may 
not be adequate and need to be conferred. Conversely, if the data analysis is
confirmed, then either these systems did not undergo migration at all, or this 
process was highly un-adiabatic. Either way, we can obtain important 
information of the formation process and posterior evolution of these planetary
systems. Is it thus important to stress that our aim will not be to certify 
whether the current planets are in fact in apsidal corotation, but solely if 
the orbital fits are consistent with these configurations. 

\subsection{{\it Gliese 876}}

\begin{figure}
\centerline{\includegraphics*[width=20pc]{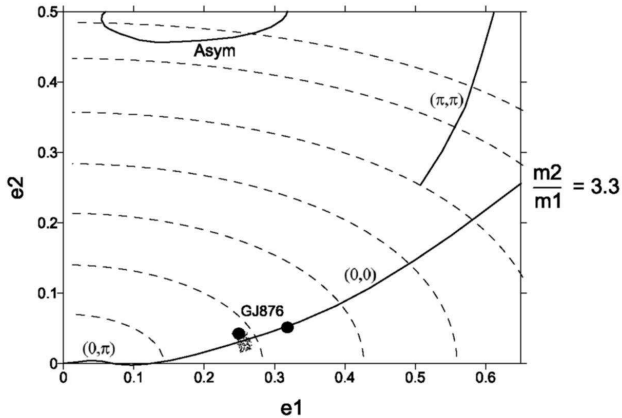}}
\caption{Corotational families for mass ratio equal to the {\it GJ876} system,
together with current orbital fits (in full circles). Dotted curves mark the
levels of constant total angular momentum.}
\label{fig10}
\end{figure}

We begin with the well known {\it GJ876} planets. Current information for this 
system gives two different possible orbits (see Laughlin and Chambers 2001, 
Lee and Peale 2002), depending on the chosen observational data base. 
Dynamical fits using Keck+Lick observations yield $(e_1,e_2)=(0.27,0.10)$, 
while data from Keck alone gives $(e_1,e_2)=(0.33,0.05)$. For both, however, 
the mass ratio is very similar, $m_2/m_1 \sim 3$. Figure \ref{fig10} shows the 
different families of corotations for $m_2/m_1=3.3$ in the eccentricity plane. 
The real bodies are represented by filled circles. Note that both 
observational fits lie very close to the zero-amplitude solutions, and it is 
easy to deduce their evolutionary track from initially circular orbits. Thus, 
we can know that the planets were initially captured in an anti-aligned 
corotation, but switched to an aligned orbit when $e_1$ surpassed the critical 
value $e_c = 0.1$. 

Together with the present orbits, we have also plotted (in small dots) the
temporal variation of the eccentricities during a $10^5$ timespan. The Keck
orbit is very close to the actual zero-amplitude corotation, so the dots are
masked within the filled circle. The other orbit, however, shows a perceptible
oscillation around the corotational family. This behavior can in fact be
predicted from the invariance of the total angular momentum. Writing this
explicitly from equation (\ref{eq4}) and supposing that the magnitude in the
temporal variation of the eccentricity is much larger than in semimajor axis, 
we find that the level curves of constant $J_{tot}$ in the eccentricity plane
are given by the expression:
\be
\label{eq12}
L_1 \sqrt{1-e_1^2} + L_2 \sqrt{1-e_2^2} =const.
\ee
where the Delaunay momenta $L_i$ can be fixed at exact resonance (see Zhou et 
al. 2004 for a similar analysis). The broken curves in Figure \ref{fig10} show 
the level curves of this integral, for different values of the constant. Note 
that the variation of the eccentricities for the Keck+Lick data follows 
closely their trend.

\subsection{{\it HD82943}}

Although not many new results were obtained from this previous system, it is 
useful as an example where the adiabatic migration scenario yields
results consistent with the observational orbital fits. The {\it HD82943} 
planets, however, are much more compromising.

\begin{figure}
\centerline{\includegraphics*[width=20pc]{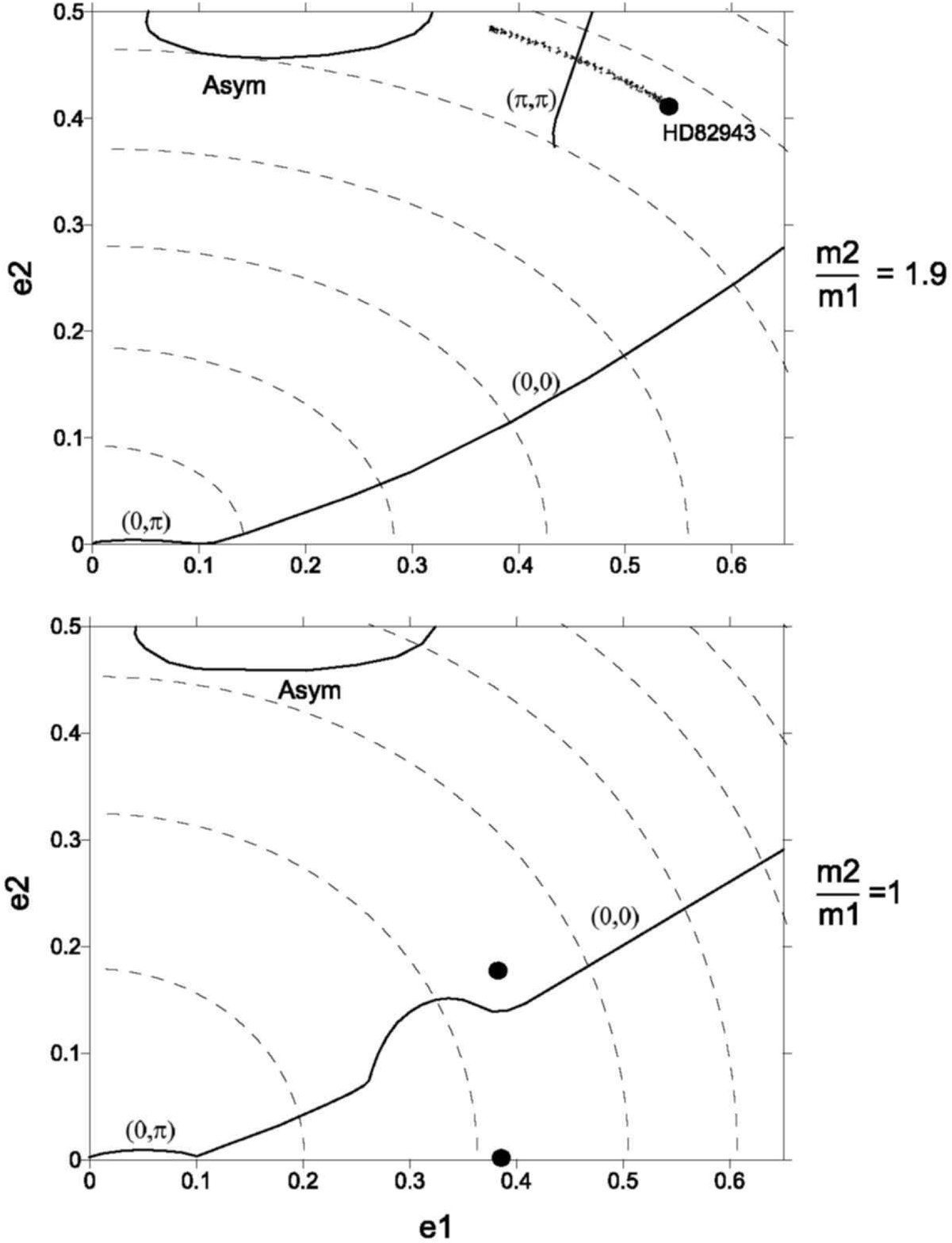}}
\caption{Same as previous figure, but for the {\it HD82943} planets. Top 
graph corresponds to $m_2/m_1=1.9$ and eccentricities from data fits published
in the Exoplanet encyclopedia homepage. Bottom: $m_2/m_1=1$ and orbital fits
from Mayor et al (2004). Note: Currently, there is a general agreement that
the orbital fit shown in the top graph is not correct. Thus, these results
represent a ``fictitious'' system.}
\label{fig11}
\end{figure}

Until very recently, the best observational data for this system was consistent
with $m_2/m_1 = 1.9$ and eccentricities $(e_1,e_2)=(0.54,0.41)$ (see 
Extrasolar Planets Encyclopedia homepage: www.obspm.fr/planets). Ji et al. 
(2003) showed that this configuration is only stable if both planets 
are trapped in a $(\pi,\pi)$-corotation. Figure \ref{fig11}a shows all the 
families of corotation for this mass ratio. Once again, the orbits fits are 
presented by a filled circle, and the levels of constant angular momentum by 
broken curves. Via a numerical integration, the temporal variation of the 
eccentricities are shown with small dots. We note a large-amplitude oscillation
around the corotational family, although this system seems to be very stable 
over large timescales.

However, a problem arises when we try to deduce the evolutionary track of
these planets from initially circular orbits. After performing numerous
numerical integrations, we could not find a single initial condition or choice
of parameters for the dissipative force that allowed a jump from the
family of aligned corotations to the $(\pi,\pi)$ case. Both families are not
only disconnected, but are separated by a region not protected from collisions;
thus there seems to be no road from one to the other that does not lead to a 
physical disruption of both planets. 

In a recent communication, Lee and Peale (2003) argued that this orbital fit
is not consistent with a smooth planetary migration, unless {\it (i)} the
planets suffered a significant mass variation during the migration or, 
{\it (ii)} the orbits are not coplanar. The first alternative seems unlikely,
since the collision curve does not depend on the masses. Thus, unless the 
planets were virtually insignificant at the time of the orbit intersection,
they would still have suffered a physical encounter leading to a disruption
or an ejection. The existence of a mutual inclination sufficiently large to
avoid close encounters is also questionable, since there is no indication
from cosmogonic theories that massive planets could form at highly non-planar
orbits. Migration simulations by Thommes and Lissauer (2003) also show no
inclination excitation for eccentricities below $~0.65$. The authors finally 
considered a different option, imagining that the capture into this type of 
corotation was obtained as a consequence of a close encounter of one of the 
bodies with a third planet. As a result, this other body was ejected and 
engulfed by the star. This alternative found certain support in new spectral 
analysis of the {\it HD82943} star showing evidence of $Li^6$ (Israelian et al.
2001), consistent with the recent absorption of a planet. 

Two other alternatives also exist: either the planets are not in apsidal 
corotation at all (thus questioning the planetary migration scenario), or the 
orbital fit is not correct. A recent paper by Mayor et al. (2004) seems to 
indicate that the latter option may be true. The authors presented a new 
orbital fit, which yields $m_2/m_1 \simeq 1$ and eccentricities 
$(e_1,e_2)=(0.38,0.18)$, thus significantly different from the previously 
published values. The same paper also shows a second new orbital fit, with the 
same mass ratio but $(e_1,e_2)=(0.38,0.0)$. From their data analysis, both 
fits have similar residues, although there seems to be a marginal preference 
for the first.

Figure \ref{fig11}b shows our analysis of these new data. Note that for 
$m_2/m_1 = 1$ there is no $(\pi,\pi)$ family, and the lower curve now has a
hump corresponding to asymmetric solutions. Once again, both orbital sets
are shown as filled circles. The first thing we note is that both fits are
much more consistent with apsidal corotations than the results shown in the top
graph. However, the solution $(e_1,e_2)=(0.38,0.0)$ seems to be rather
distant from its zero-amplitude curve. A simple analysis of the levels of
constant $J_{tot}$ shows that, if this is an actual apsidal corotation, the 
temporal variation of the eccentricities would be very large. The 
$(e_1,e_2)=(0.38,0.18)$ fit, however, is fairly close to the 
zero-amplitude curve, thus much more likely. However, in order to complete the
analysis, we must also consider the values of the angular variables. From 
Figure \ref{fig2} we obtain that a small-amplitude corotation for this point
should have $\Delta \varpi \sim 50$ degrees. The actual value of this angle is
$110$ degrees, thus also indicating a large-amplitude apsidal corotation in 
the best case scenario.

\begin{figure}
\centerline{\includegraphics*[width=20pc]{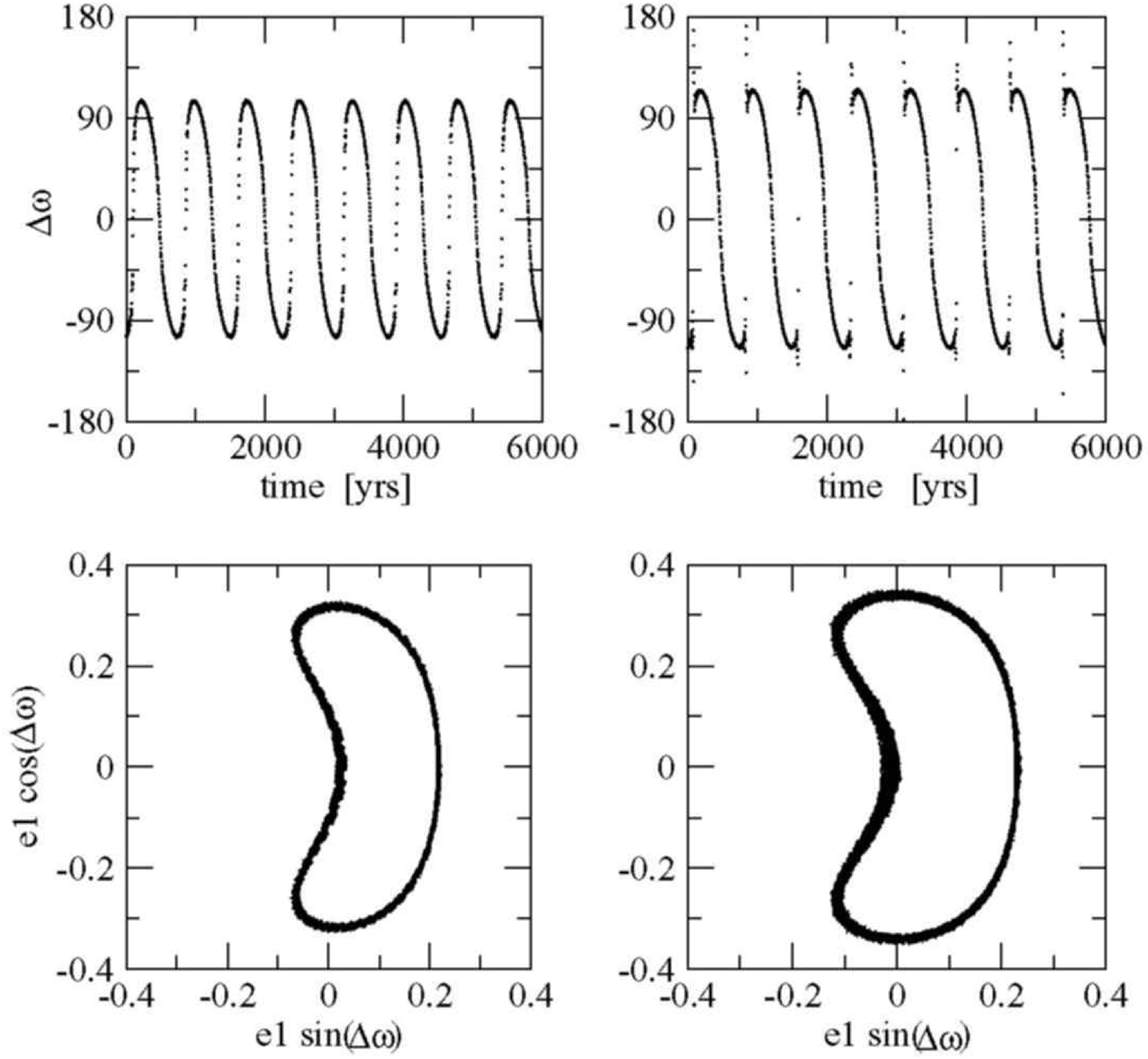}}
\caption{Numerical simulation of two stable initial conditions in the vicinity
of the Mayor et al (2004) data for the {\it HD82943} planets. Left-hand plots
show a corotation, while the right-hand a paradoxic $\theta_1$-libration
with a circulation of the difference of pericenter. See text for details.}
\label{fig12}
\end{figure}

In order to perform a more detailed study, we considered the Mayor et al. 
(2004) data for the planets including the uncertainties in each orbital 
element. We then generated a total of 100 initial conditions for the two 
planets according to a homogeneous distribution of orbital elements inside 
the error bars. Each resulting fictitious system was then integrated 
numerically for one million years. We found that 80\% were unstable, ending in 
physical collisions or escape of one of the masses. Of the remaining, 15 were 
found in stable large-amplitude apsidal corotations (simultaneous libration 
$\theta_1$ and $\Delta \varpi$), while 5 of the initial conditions yielded 
apparent $\theta_1$-librations but with a circulation of the difference in 
longitudes of pericenter. Typical results are shown in Figure \ref{fig12}, 
where the left-hand plots correspond to a apsidal corotation and the 
right-hand shows one of the $\theta_1$-librations. Top graphs present the 
temporal variation of $\Delta \varpi$ for a time interval of 6000 years, while 
the lower graphs show the orbits in the regular variables 
$(e_1 \cos{\Delta \varpi}, e_1 \sin{\Delta \varpi}$. We can clearly see that
although the angle does circulate in the right-hand side, this is really 
topologically equivalent to the corotational solution. Moreover, the geometric
average of the eccentricities in both solutions yield the same values
$(e_1,e_2) = (0.35,0.17)$ , which is very close to the corotational family
shown in Figure \ref{fig11}.

As a conclusion of this system, we see that sometimes a dynamical analysis 
of the present planetary orbits is not enough to ascertain that a given 
orbital determination is consistent with planetary migration. The evolutionary
tracks can yield important information, and help identify problematic cases. 
Once noted, then we can study whether the problem arises from orbital 
uncertainties, or if they point towards real dynamical evolution. Furthermore,
the new orbital fit of this system is completely compatible with corotational
solutions and, thus, with a planetary migration scenario.

\subsection{{\it HD160691}}

From the previous discussions, we can now divide the eccentricity plane into
three distinct regions. These can be deduced from Figure \ref{fig1}. First, 
the NS (No Solution) region marks time-averaged values of the orbital 
eccentricities which are not allowed, unless the bodies are not in apsidal 
corotation, independent of any past migration. A second region groups the 
$(\pi,\pi)$ and high-eccentricity asymmetric corotations. These solutions are 
not joined to quasi-circular orbits through continuous families. Thus, the 
presence of any real planetary system would be indication of a problematic 
case. It may indicate a catastrophic past (e.g. close encounters, ejection of 
missing planets, etc.) or be evidence that the planets are not in apsidal 
corotation at all. Once again, this would then be good evidence against 
planetary migration, at least in an adiabatic process. Finally, a third region 
(including aligned, anti-aligned and asymmetric apsidal corotations) shows 
those orbits whose evolutionary tracks are consistent with adiabatic migration.

As an example of the application of this plot, we analyze the {\it HD160691}
system. According to (Jones et al. 2002), two planets orbit this star, with
$m_2/m_1 = 0.6$, and $(e_1,e_2)=(0.31,0.80)$. Dynamical analysis by Bois et al.
(2003) confirms the long term stability of this fit in an apsidal corotation. 
Nevertheless, a look at Figure \ref{fig12} and the mass level curves of Figure 
\ref{fig3}, shows that even if this orbital fit is consistent with a 
corotational solution, the evolutionary track for this mass ratio cannot 
explain these present eccentricities. Thus, once again, and similarly with the 
previous system, we have found a problematic case. However, (and once again), 
recent observations (Mayor et al., private communication) and new orbital fits 
gave raised severe doubts as to the actual existence of the {\it HD160691}c 
planet. Thus, it seems that this resonant system is probably just as artifact
of insufficient observational data.

\section{Conclusions}

In this paper we have presented a new catalog of general corotational solutions
for the 2/1 mean-motion resonance. Apart from the well-known aligned and
anti-aligned solutions, we have also extended our knowledge of asymmetric
configurations and have mapped the recently discovered $(\pi,\pi)$-corotations.
Since these periodic orbits depend on the planetary masses only through the
ratio $m_2/m_1$, and on the semimajor axes only by $a_1/a_2$, they are very
general in nature and should be valid for any exoplanetary system showing
two planets trapped in this commensurability. 

The determination of the period of oscillation $\tau_c$ around these fixed 
points of the averaged problem shows that any migration mechanism with 
characteristic timescale satisfying  
\be
\label{eq13}
{1 \over a_i} {da_i \over dt} \gg {m_1 m_2 \over a_i} \tau_c 
\ee
should be adiabatic. Thus, starting from quasi-circular orbits, the 
evolutionary track of the planets inside the resonance should be well 
reproduced by the families of apsidal corotations for that particular mass 
ratio. This, together with the fitted values of the orbital eccentricities, 
allow us to stipulate whether present orbits are consistent with apsidal 
corotations and, consequently, with a planetary migration of the system from 
cosmogonic locations far from the star. In other words, we are able to suggest 
a simple test which relates the present orbits with restrictions on the 
properties of the formation process of resonant planets. 

Application to the 2/1 resonance shows that, of the three published systems,
{\it GJ876} satisfies all our conditions satisfactorily. The old orbits 
for the {\it HD82943} planets are not consistent with our hypothetical
relation between corotation and migration. However, new orbital determinations
for this system have yielded values very compatible with orbital decay, thus 
indicating that the problem is due to uncertainties in the fits themselves. 
Finally another problematic case, {\it HD160691}, also seems to be biased by 
errors in the orbital fits. In fact, the exterior planet of this system is
probably not existent at all.

In view of these examples, it seems that there is not indication of the
existence of any planetary system in the 2/1 commensurability which is in
specific contradiction with the hypothesis of planetary migration. Recent
results by Zhou et al. (2004) for the {\it 55 Cnc} system trapped in the 3/1
mean-motion resonance show similar results. It will be very interesting to 
extend this analysis to all future resonant systems and see whether an 
adiabatic migration mechanism continues to pass all the tests.

\section*{Acknowledgments}
This work has been supported by the Argentinian Research Council -CONICET-, 
the Brazilian National Research Council -CNPq- through the fellowship 
300946/96-1, as well as the S\~ao Paulo State Science Foundation FAPESP.

\end{document}